\documentclass[conference, review]{IEEEtran}
\IEEEoverridecommandlockouts
\usepackage{cite}
\usepackage{amsmath,amssymb,amsfonts}
\usepackage{graphicx}
\usepackage{textcomp}
\usepackage{xcolor}
\usepackage{orcidlink}

\usepackage{tikz}

\usepackage{algorithmicx}
\usepackage[ruled]{algorithm}
\usepackage[noend]{algpseudocode}

\usepackage{multirow}
\usepackage{booktabs}
\usepackage{pifont}
\newcommand{\cmark}{\ding{51}}%
\newcommand{\xmark}{\ding{55}}%
\usepackage{subcaption} 

\usepackage[colorinlistoftodos,prependcaption,textsize=small]{todonotes} 

\newcommand{\Albatross}{Aquarius}

\newcommand{\eg}{\textit{e.g.},\ }

\newcommand{\publicationtype}{paper}

\DeclareMathOperator*{\argmin}{arg\,min}

\def\BibTeX{{\rm B\kern-.05em{\sc i\kern-.025em b}\kern-.08em
    T\kern-.1667em\lower.7ex\hbox{E}\kern-.125emX}}
\begin{document}

\title{Efficient Data-Driven Network Functions}
\author{Zhiyuan Yao$^{\orcidlink{0000-0002-7211-1506}}$, Yoann Desmouceaux$^{\orcidlink{0000-0001-6322-2338}}$, Juan-Antonio Cordero-Fuertes$^{\orcidlink{0000-0001-5771-3122}}$, Mark Townsley$^{\orcidlink{0000-0001-7976-3470}}$, Thomas Clausen$^{\orcidlink{0000-0002-7400-8887}}$
\thanks{Z. Yao, Y. Desmouceaux and M. Townsley are with Cisco Systems Paris Innovation and Research Laboratory (PIRL), 92782 Issy-les-Moulineaux, France; emails \{yzhiyuan,ydesmouc,townsley\}@cisco.com.}
\thanks{Z. Yao, J.-A. Cordero-Fuertes and T. Clausen are with \'Ecole Polytechnique, 91128 Palaiseau, France; emails \{zhiyuan.yao,juan- antonio.cordero-fuertes,thomas.clausen\}@polytechnique.edu.}
\thanks{This work is, in part, supported by the Cisco endowed ``Internet Technologies and Engineering'' chaire at \'Ecole Polytechnique.}
}

\maketitle

\begin{abstract}
    Cloud environments require dynamic and adaptive networking policies.
    It is preferred to use heuristics over advanced learning algorithms in Virtual Network Functions (VNFs) in production becuase of high-performance constraints.
    This paper proposes Aquarius to passively yet efficiently gather observations and enable the use of machine learning to collect, infer, and supply accurate networking state information – without incurring additional signalling and management overhead. 
    This paper illustrates the use of Aquarius with a traffic classifier, an auto-scaling system, and a load balancer – and demonstrates the use of three different machine learning paradigms – unsupervised, supervised, and reinforcement learning, within Aquarius, for inferring network state. 
    Testbed evaluations show that Aquarius increases network state visibility and brings notable performance gains with low overhead.
\end{abstract}


\section{Introduction}
\label{sec:intro}

To increase network programmability, and balance the trade-off between capital expenditures and quality of service (QoS), Virtual Network Functions (VNFs) (\eg firewalls, load balancers) replace or augment dedicated hardware devices and play a significant role in large-scale data centers (DCs), running on commodity computing platforms.
To dynamically monitor and configure VNFs, the routing and decision-making process (\textit{control plane}) is dissociated from the network packets forwarding process (\textit{data plane}).

Data-driven mechanisms based on machine learning (ML) and reinforcement learning (RL) algorithms are applied in the control plane to adaptively manage networking policies~\cite{decima2018, auto2018sigcomm, yao2021reinforced}.
For instance, auto-scaling systems and load balancers can achieve improved QoS with reduced cost based on periodically polled resource utilisation of distributed nodes~\cite{smartsla}.
However, it is challenging to apply these algorithms in networking systems in real-time, since they require fine-grained observations of network and system states~\cite{sivakumar2019mvfst}.

Reactive polling resource utilisation and system performance incur additional control messages~\cite{lbas-2020, 6lb, smartsla, fu2021use} and reduce system scalability.
Fine-grained networking features are extracted offline or in simulated environments for clustering and RL algorithm development~\cite{yang2018elastic,sivakumar2019mvfst}.
Since the data plane is constrained by low-latency and high-throughput requirements~\cite{vpp}, heuristics--which may not be adaptive to dynamic environments--prevail over advanced learning alrogithms in real-world high performance networks~\cite{aws-elastic, maglev, silkroad2017, 6lb, hpcc, fu2021use}.

This \publicationtype\ proposes \Albatross, a fast and scalable data collection and exploitation mechanism that bridges different requirements for data planes (low-latency and high-throughput) and control planes (making informed decisions).
Extensive performance and overhead evaluations of \Albatross\ in a realistic testbed show that Aquarius enable:
\begin{itemize}
	\itemsep-0.1em 
	\item \textbf{unsupervised learning $+$ offline data analysis}: 
	creating benchmark datasets to gain insight in different networking problems with minimal data collection overhead;
	\item \textbf{supervised learning $+$ VNF management}:
	embedding ML techniques to achieve self-aware monitoring and self-adaptive orchestration in an elastic compute cloud;
	\item \textbf{reinforcement learning $+$ online policy updates}:
	enabling closed-loop control to optimise routing policies and improve QoS, with no human intervention.
\end{itemize}


\begin{table*}[t!]
    \begin{center} 
        \caption{Comparison of data-driven VNF systems.}
	    \label{tab:background-related}
        \resizebox{\textwidth}{!}{ 
            \begin{tabular}{l|c|c|c|c|c|c|c|c|c|c|c}
            \toprule
            \textbf{Property$\slash$Work}  & \cite{opnfv2019, openstack} & \cite{le2017uno,aws-nitro,bari2013policycop} & \cite{gonccalves2009evaluation} & \cite{smartsla, chen2018survey} & \cite{sivakumar2019mvfst, 6lb, lbas-2020, decima2018} & \cite{clove, auto2018sigcomm} & \cite{tuor2017deep} & \cite{lalitha2016traffic} & \cite{xiong2019switches} & \cite{taurus2020} & \textbf{\textit{\Albatross}} \\ \midrule
            \textbf{No Control Message}         & \xmark & \xmark & \xmark & \xmark & \xmark & \cmark & \cmark & \xmark & \cmark & \cmark & \cmark \\
            \textbf{Distributed}                & \xmark & \xmark & \xmark & \xmark & \cmark & \cmark & \cmark & \cmark & \cmark & \cmark & \cmark \\
            \textbf{Commodity Device}           & \cmark & \xmark & \cmark & \cmark & \cmark & \cmark & \cmark & \cmark & \xmark & \xmark & \cmark \\
            \textbf{Functionality}              & Framework & \multicolumn{2}{c|}{Management Protocol} & Autoscaling & \multicolumn{2}{c|}{Traffic Optimisation} & \multicolumn{3}{c|}{Traffic Classification} & Framework & Framework \\
            \bottomrule
            \end{tabular}
        }
	\end{center}
    \vskip -0.25in
\end{table*}

\section{Background}
\label{sec:background}

This section presents the challenges of effective feature collection and data-driven VNFs in cloud DCs, and, with a comparison of related work, motivates the design of \Albatross.

\subsection{Challenges}
\label{sec:background-challenge}

There is a rising trend of embedding intelligence and applying ML techniques in cloud and distributed systems to dynamically monitor and adaptively configure system parameters and characteristics (\eg server configurations, forwarding rules)~\cite{survey-anomaly, tuor2017deep, sivakumar2019mvfst, fu2021use}.
However, this raises a number of challenges and trade-offs:

\textbf{Online and Reliable Feature Collection:}
Few reliable datasets are available and considered as benchmark for ML applications in VNFs and networking systems (\eg traffic analysis and anomaly detection)~\cite{kdd-issues, predict, nsl-kdd, caida}.
Though log-based feature collection provides abundant information for various types of applications, it incurs high overhead under heavy traffic, which leads to inaccurate and unreliable measurements and makes it hard to bring ML algorithms ``online'' (making inference in real time)~\cite{survey-anomaly}.

\textbf{Scalability vs. Visibility:} 
Active probing is another way of feature collection for VNFs to monitor the system state and make informed decisions~\cite{opnfv2019, openstack, lbas-2020, lalitha2016traffic}.
However, this incurs additional communication overhead and requires modifications on each node to maintain management and communication channels.

\subsection{Requirements}
\label{sec:background-requirements}

Based on the challenges, this \publicationtype\ summarizes the following requirements to enable data-driven VNFs in the cloud:

\textbf{Universality:} the feature collection mechanism should cover a wide range of features and be application-agnostic;

\textbf{Reliability:} the collected features should be representative, and have high usability and granularity;

\textbf{Scalability:} the feature collection and exploitation mechanism should incur minimal performance overhead and support large-scale and dynamically changing network topology;

\textbf{Deployability:} the mechanism should be plug-and-play and require no additional installation or configuration.

\begin{figure}[t]
	\centering
	\includegraphics[width=.9\columnwidth]{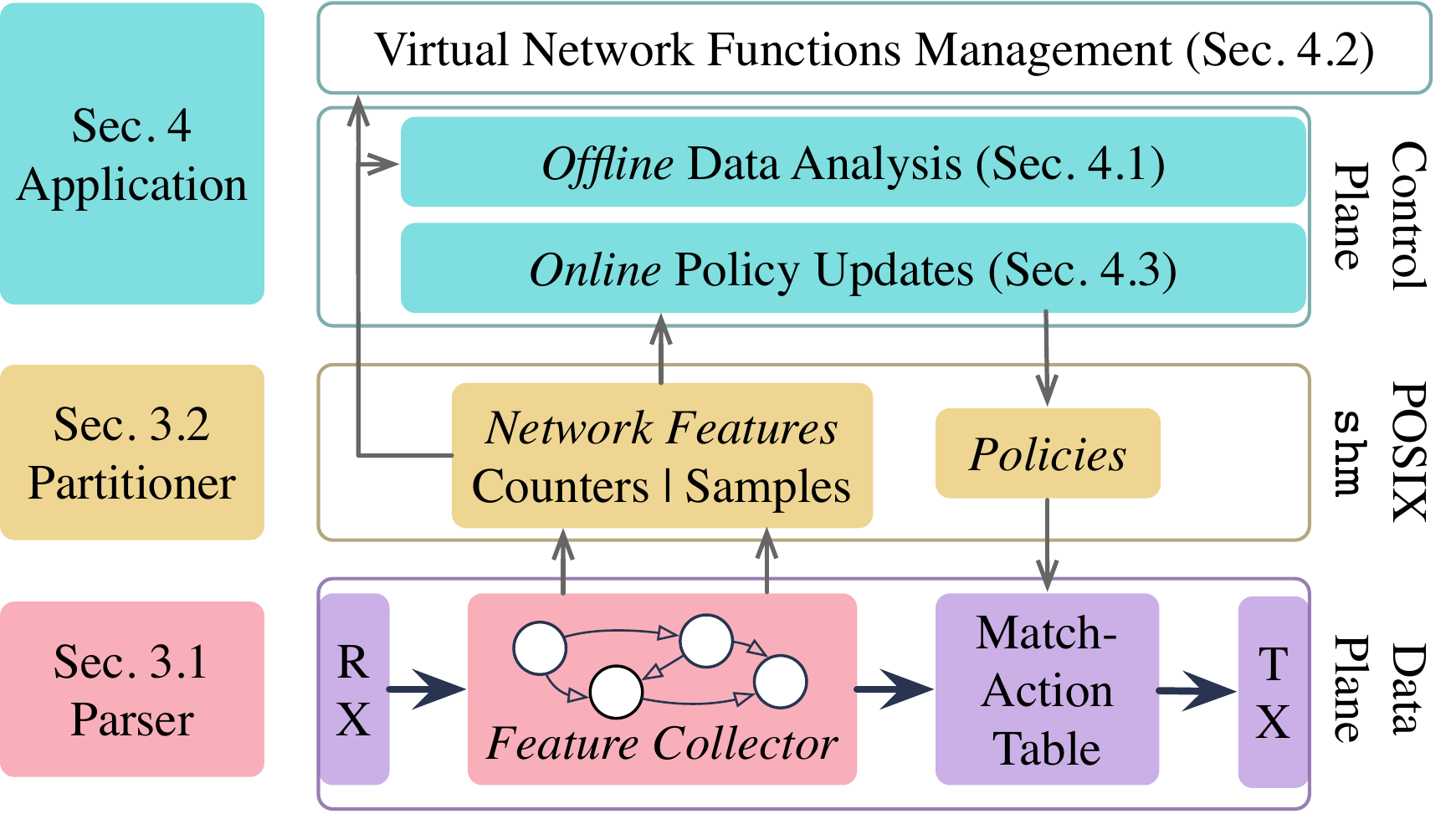}
	\caption{\Albatross\ architecture overview.}
	\label{fig:design-overview}
	\vskip -0.1in
\end{figure}

\subsection{Related Work}
\label{sec:background-related}

Various mechanisms (summarised in Table~\ref{tab:background-related}) dynamically configure and manage VNFs, making data-driven decisions.

ML allows inferring system states from networking features, for, \eg intrusion detection systems~\cite{tuor2017deep}, traffic classification~\cite{nguyen2008survey, pacheco2018towards}, and task scheduling~\cite{auto2018sigcomm, decima2018}.
To obtain networking features, these ML applications operate at the Application Layer.
However, they are not application-agnostic and do not generalise to different use cases.
\Albatross\ collects a wide range of features at the Transport Layer and enables generic data-driven network functions with minimal overhead.

Management and Orchestration (MANO) frameworks use centralised controllers to monitor and update VNF configurations~\cite{opnfv2019, openstack}.
Software-Defined Network (SDN) provides programmable APIs to gather per-flow or application-level features in a centralised way, to adaptively update configurations, using network equipments that supports the OpenFlow protocol~\cite{bari2013policycop}.
Smart Network Interface Cards (sNICs)~\cite{le2017uno} and Nitro~\cite{aws-nitro} offload VNFs from host processors to dedicated hardware devices to boost performance and reduce operational cost with centralised management.
\Albatross\ passively extracts networking features from the data plane and let VNFs make decisions in a distributed way.

Distributed VNFs also benefit from periodically polled network states (\eg CPU and memory usage), to ensure service availability, improve QoS~\cite{lbas-2020}, or classify networking traffic~\cite{lalitha2016traffic}.
Some network functions gain more visibility via in-network telemetry (INT)~\cite{clove}.
However, these methods require to either deploy agents, or to modify protocol stack on network nodes, which reduce the deployability.
\Albatross\ employs the plug-and-play design and requires no coordinated modification in the network.

Learning algorithms incurs additional inference and processing latencies.
To reduce latency, dedicated hardware, \eg CGRA~\cite{taurus2020}, and NetFPGA~\cite{xiong2019switches}, helps accelerate data processing for in-network ML applications.
Yet they lack flexibility when developing ML algorithms for different use cases in elastic networking systems.
MVFST-RL~\cite{sivakumar2019mvfst} proposes--in simulators--to asynchronously update networking configurations from learning algorithms to reduce additional latency in the data plane.
\Albatross\ can incorperate intelligence in a variety of VNFs, requiring no dedicated device, yet it is ready to be deployed in real-world systems.



\section{Design}
\label{sec:design}

To meet the $4$ requirements summarised in Section~\ref{sec:background-requirements}, \Albatross\ is designed as a $3$-layer architecture (Fig.~\ref{fig:design-overview}).
\Albatross\ embeds a feature collector at the Transport Layer in the data plane, to efficiently and passively extracts a wide range of features with high granularity and low performance overhead.
It makes the features available via shared memory, for applications of ML algorithms on various use cases.
This \publicationtype\ illustrates the design using TCP traffic, which prevails in the cloud of Content Delivery Networks (CDNs)~\cite{facebook-dc-traffic}.

\subsection{Parser Layer}
\label{sec:design-parser}

To balance the tradeoff between scalability and visibility, networking features which indicate system states can be passively collected from the data plane to avoid active probing and additional installations and configurations.

\begin{figure}[t]
	\centering
	\includegraphics[width=.9\columnwidth]{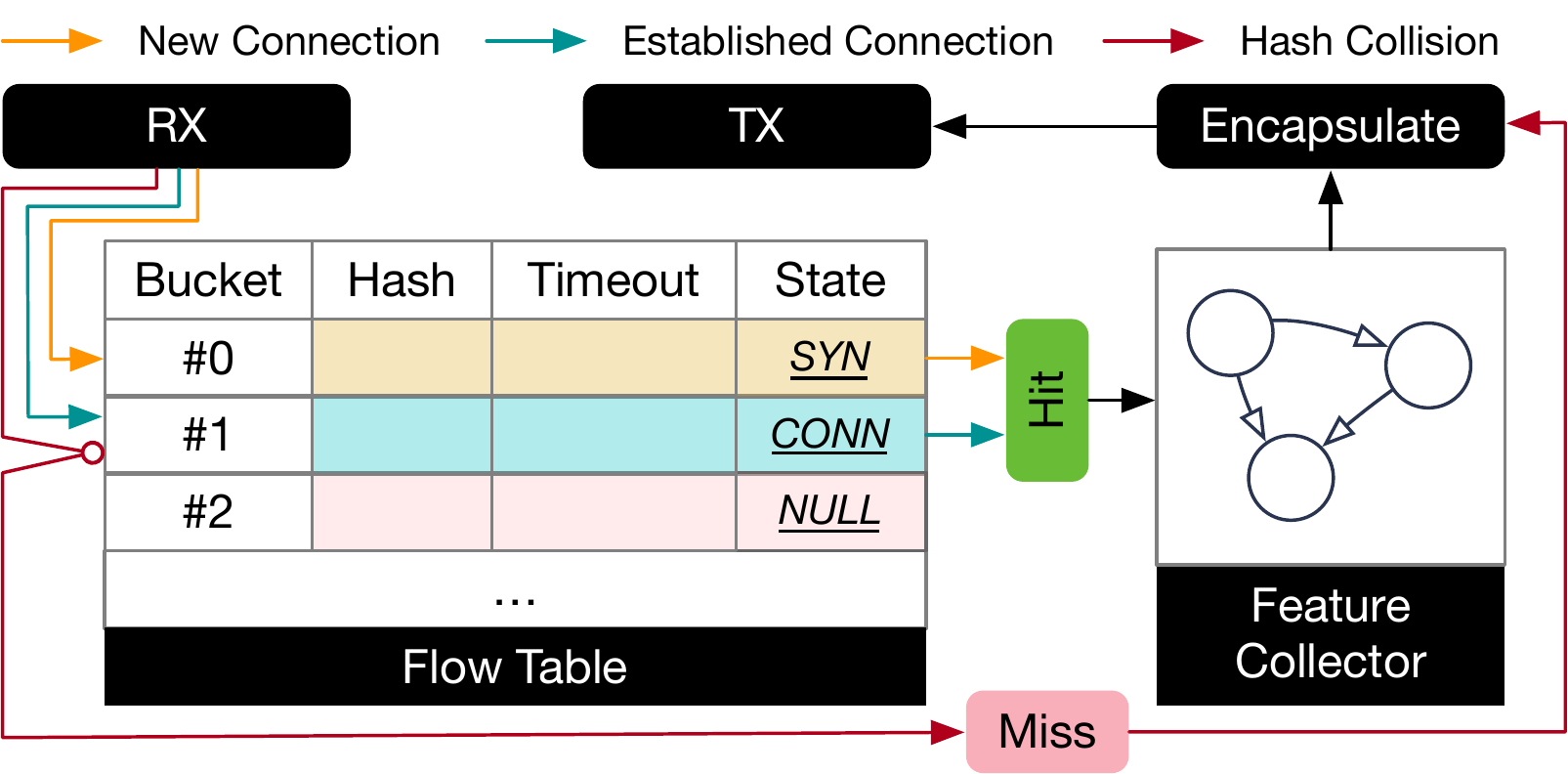}
	\caption{Flow table data structure and workflow.}
	\label{fig:feature-flow-table}
	\vskip -0.1in
\end{figure}

\begin{figure}[t]
	\centering
	\includegraphics[width=.9\columnwidth]{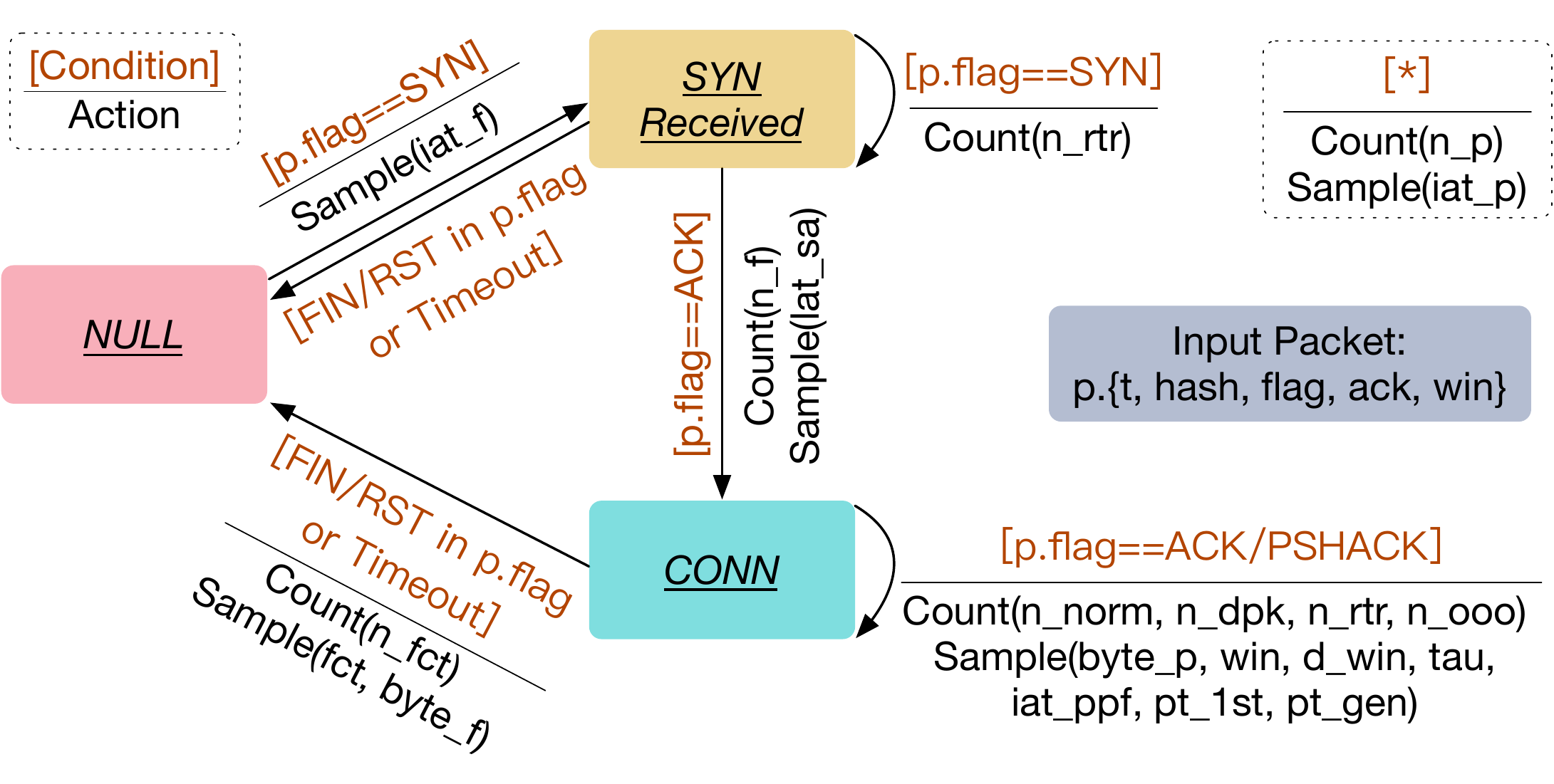}
	\vskip -0.05in
	\caption{A state machine of feature collector for TCP traffic.}
	\label{fig:feature-state-machine}
	\vskip -0.1in
\end{figure}

\subsubsection{Stateful Feature Collection}
\label{sec:design-parser-collect}

Network traffic consists of flows that traverse different nodes (\eg edge routers, load balancers, servers) in the system, whose states can be traced and retrieved from the flows--along with traffic characteristics.

\textbf{Rationale:} 
Stateless feature collection mechanisms (\eg sketches~\cite{yang2018elastic}) do not track the state of network flows, yet they can gather counters as ordinal features for ML algorithms using hashing functions, with little performance overhead.
However, ordinal features contains less information than quantitative features--time-related features (\eg round-trip time, inter-arrival time, flow duration) and throughput information (\eg congestion window size, flow size), which are not captured by stateless mechanisms.

\textbf{Design:}
\Albatross\ tracks flow states in bucket entries with a stateful table (Fig.~\ref{fig:feature-flow-table}), which can be configured to collect a wide range of features using a state machine depicted in Fig.~\ref{fig:feature-state-machine}.
In the flow table, \Albatross\ stores the information of each flow into a bucket entry indexed by $hash(fid)\%M$, where $fid$ is the flow ID (\eg the $5$-tuple of TCP flows) and $M$ is the flow table size.
An entry in the flow table can be in one of three states--\texttt{SYN}, \texttt{CONN} and \texttt{NULL} (Fig.~\ref{fig:feature-state-machine}).
The flow state transitions trigger the networking feature updates, which are described in the next subsection.
In case where the bucket entry is not available when a new flow arrives, the flow is considered as a ``miss'' and is excluded by the feature collector.

\subsubsection{Network Features}
\label{sec:design-parser-categorize}


Various features can gainfully benefit decision making process for different use cases.

\textbf{Rationale:}
As a generic feature collection mechanism, \Albatross\ should be able to collect as much information as possible with minimal overhead (\eg memory space consumption).
Other than ordinal and quantitative features, to capture the system dynamic, it is also important to trace the timestamps of different events, for sequential ML algorithms.

\textbf{Design:}
With the flow table, \Albatross\ allows flexible configuration of attributes, to gather the most significant features and optimise the memory usage overhead for different applications.
Quantitative features are collected as samples, using reservoir sampling (Algorithm~\ref{alg:feature-reservoir}).
Since networking environments are dynamic, it is important to capture not only the features, but also the sequential information of the system.
Reservoir sampling gathers a representative group of samples in fix-sized buffer from a stream.
It captures both the sampling timestamps and exponentially-distributed numbers of samples over a time window, which help analyse patterns sequentially.

\alglanguage{pseudocode}
\begin{algorithm}[t]
\footnotesize
\caption{Reservoir sampling with no rejection}
\label{alg:feature-reservoir}
\begin{algorithmic}[1]
\State $k \gets$ reservoir buffer size
\State $buf \gets [(0, 0), \dots, (0, 0)]$\Comment{Size of $k$}
\For {each observed sample $v$ arriving at $t$}
	\State $randomId \gets rand()$
	\State $idx \gets randomId\%N$ \Comment{randomly select one index}
	\State $buf[idx] \gets (t, v)$ \Comment{register sample in buffer}
\EndFor
\Statex
\end{algorithmic}
  \vspace{-0.4cm}%
\end{algorithm}

\begin{figure}[t]
	\centering
	\includegraphics[width=.9\columnwidth]{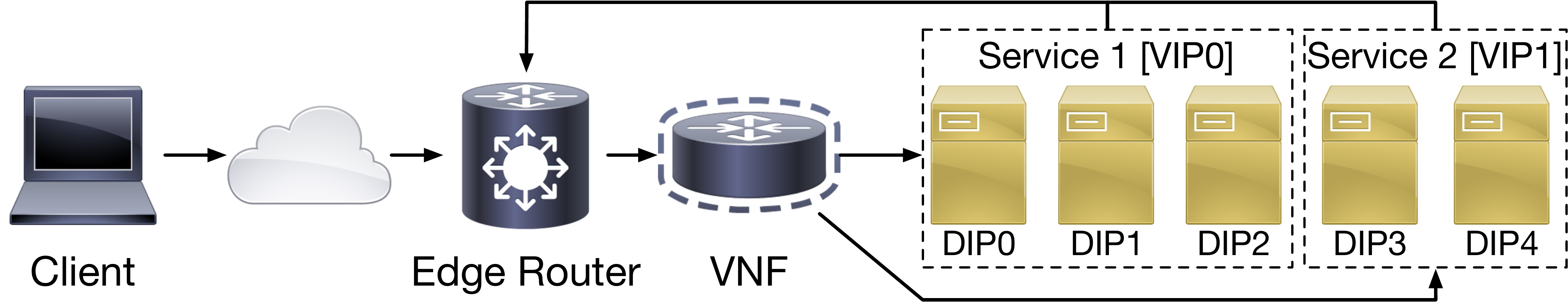}
	\caption{Cloud service topology.}
	\label{fig:feature-dsr}
	\vskip -0.1in
\end{figure}

\subsection{Partitioner Layer}
\label{sec:design-partitioner}

Cloud services have different characteristics and they are identified by virtual IPs (VIPs) (Fig.~\ref{fig:feature-dsr}), corresponding to clusters of provisioned resources--\eg servers, identified by unique direct IPs (DIPs).
In production, cloud DCs are subject to high traffic rates and their environments and topologies change dynamically.

\textbf{Rationale:}
Different cloud services should be separated to (i) avoid multimodal distributions in collected features and to (ii) allow dynamically adding or removing services.
Features should be made available so that both spatial and sequential information can be easily partitioned and accessed.
Even with heavy traffic and dynamically changing network topology, features should be reliable and easy to access with low latency.

\begin{figure}[t]
	\begin{center}
		\centerline{\includegraphics[width=\columnwidth]{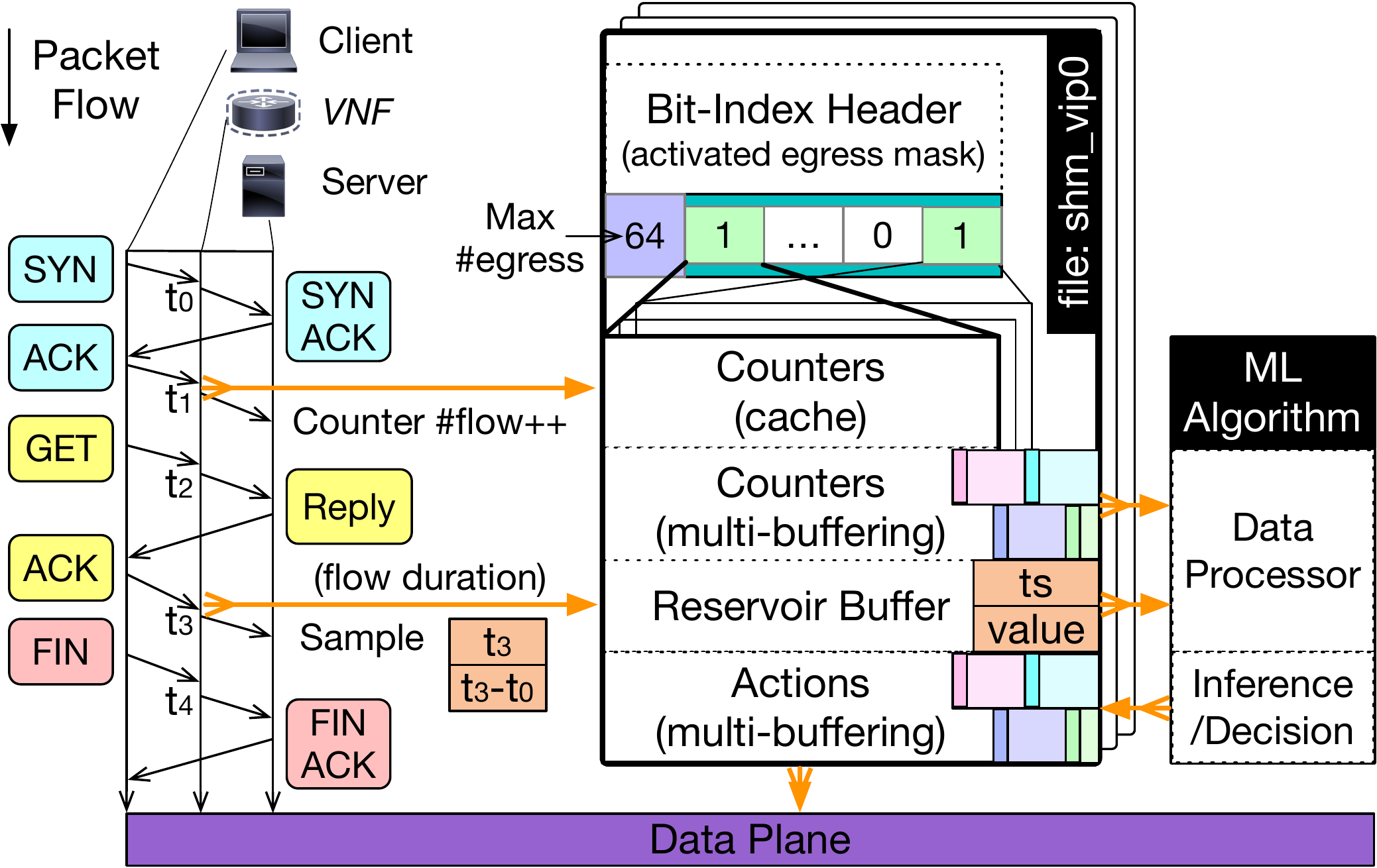}}
		\caption{\Albatross\ \texttt{shm} layout and data flow pipeline.}
		\label{fig:architecture-mechanism}
	\end{center}
	\vskip -0.2in
\end{figure}

\textbf{Design:}
\Albatross\ organises observations of each VIP in independent POSIX shared memory (\texttt{shm}) files, to provide scalable and dynamic service management.
In each \texttt{shm} file, collected features are further partitioned by egress equipments so that spatial information can be distinguished.
Fig.~\ref{fig:architecture-mechanism} exemplifies the \texttt{shm} layout and data flow.

\subsubsection{Bit-Index and Masking}
\label{sec:design-partitioner-mask}

The first byte in the \texttt{shm} file of a VIP defines the max number of egress equipments $N$ ($64$ in Fig.~\ref{fig:architecture-mechanism}).
The $N$-bit \textit{bit-index header} helps quickly identify activated egress and its corresponding memory space--the $i$-th bit is set to $1$ if the $i$-th egress is active and $0$ otherwise.
Adding an activated egress to the system requires only to set the corresponding bit to $1$ after initialising its memory space.
It suffices to simply flip the bit from $1$ to $0$ to deactivate an egress node.

\begin{table}[t]
	\footnotesize
	\centering
	\begin{tabular}{|ll|c|c|}
		\hline
		\multicolumn{2}{|l|}{Operation / Complexity}                                                                                        & Computation                       & Memory               \\ \hline
		\multicolumn{2}{|l|}{Add / Remove VIP}                                                                                              & $\mathcal{O}(1)$                  & $\mathcal{O}(kN+mN)$ \\ \hline
		\multicolumn{2}{|l|}{Add egress node}                                                                                               & $\mathcal{O}(1)$                  & $\mathcal{O}(k+m)$   \\ \hline
		\multicolumn{2}{|l|}{Remove egress node}                                                                                            & $\mathcal{O}(1)$                  & $\mathcal{O}(1)$     \\ \hline
		\multicolumn{2}{|l|}{\begin{tabular}[c]{@{}l@{}}Register reservoir sample\\ Update counter (cache)\end{tabular}}                    & $\mathcal{O}(1)$                  & $\mathcal{O}(1)$     \\ \hline
		\multicolumn{2}{|l|}{\begin{tabular}[c]{@{}l@{}}Update counters / actions\\ (multi-buffering)\end{tabular}}                         & $\mathcal{O}(1)$                  & $\mathcal{O}(N)$     \\ \hline
		\multicolumn{1}{|l|}{\multirow{2}{*}{\begin{tabular}[c]{@{}l@{}}Get the latest \\observation\end{tabular}}}                                                       & 1 node    & \multirow{2}{*}{$\mathcal{O}(m)$} & $\mathcal{O}(k+m)$   \\ \cline{2-2} \cline{4-4} 
		\multicolumn{1}{|l|}{}                                                                                                  & All nodes &                                   & $\mathcal{O}(kN+mN)$ \\ \hline
		\multicolumn{1}{|l|}{\multirow{2}{*}{\begin{tabular}[c]{@{}l@{}}Update action in\\ the data plane\end{tabular}}} & 1 node    & \multirow{2}{*}{$\mathcal{O}(m)$} & $\mathcal{O}(1)$     \\ \cline{2-2} \cline{4-4} 
		\multicolumn{1}{|l|}{}                                                                                                  & All nodes &                                   & $\mathcal{O}(N)$     \\ \hline
	\end{tabular}
    \caption{Computation and memory complexity of different operations, where $k$ is the size of reservoir buffer, $N$ is the number of egress nodes, and $m$ is the level of multi-buffering.}
	\label{tab:design-complexity}
	\vskip -0.1in
\end{table}

\subsubsection{Independent Egress Memory Space}
\label{sec:design-partitioner-egress}

Each egress node has its own independent memory space, storing counters, reservoir samples, and data plane policies (actions).
As depicted in Fig.~\ref{fig:architecture-mechanism}, on receipt of the first \texttt{ACK} from the client to a specific egress node $i$, VNF increments the number of flows in the counters cache of node $i$.
Quatitative features (\eg flow duration $t_3 - t_0$ gathered at $t_3$ in Fig.~\ref{fig:architecture-mechanism}) can be stored in the reservoir buffer of node $i$ using Alg.~\ref{alg:feature-reservoir}.
Obtained features for all active egress nodes can then be aggregated and processed to make further inferences or data-driven decisions, which can be written back to the memory space of each egress node.

\subsubsection{Multi-Buffering and Asynchronous I/O}
\label{sec:design-partitioner-io}

Counters and actions are exchanged between cache and buffer using $m$-level multi-buffering with incremental sequence ID.
When copying data, the sequence ID is set to $0$ to avoid I/O conflicts.
ML algorithms can pull the latest observations and push the latest data-driven decisions using multi-buffering with no disruption in the data plane.
This design offers an asynchronous $2$-way communication interface to exchange fine-grained features and data-driven decisions with low latency.

Both computation and memory space complexity is presented in Table~\ref{tab:design-complexity}.
The whole dataflow is asynchronous and avoid stalling in the data exchange process in both the data plane and the control plane.

\subsection{Implementation}
\label{sec:micro-implement}

This \publicationtype\ implements \Albatross\ as a plugin to the Vector Packet Processor (VPP)~\cite{vpp}.
The flow table size is configured as $M=65536$.
Each sampled network feature is a $2$-tuple of a $32$-bit float timestamp and a $32$-bit value.
The reservoir buffer size is $k = 128$ for each feature per egress equipment.
The \texttt{shm} file of each VIP consists of $6$KB $3$-level multi-buffering counters and $832$KB reservoir sampling buffers.
In this \publicationtype, \Albatross\ is implemented to be able to collect $8$ ordinal features (counters) and $65$ quantitative features in total.


\section{Applications}
\label{sec:application}

This section shows $3$ applications of \Albatross\ in cloud DCs in the context of $3$ key VNFs--traffic classification, resource prediction and auto-scaling, and Layer-$4$ load balancing, along with $3$ different ML paradigms.

\subsection{Traffic Classification}
\label{sec:application-classification}

As one of the key VNFs in the cloud, traffic classification allows distinguishing different types of traffic~\cite{tuor2017deep, lalitha2016traffic,xiong2019switches}, to allocate appropriate resources and achieve service level agreements.
It also helps detect anomalies and security threats to prevent potential damages or losses~\cite{tuor2017deep}.

\begin{figure}[t]
	\centering
	\includegraphics[width=.9\columnwidth]{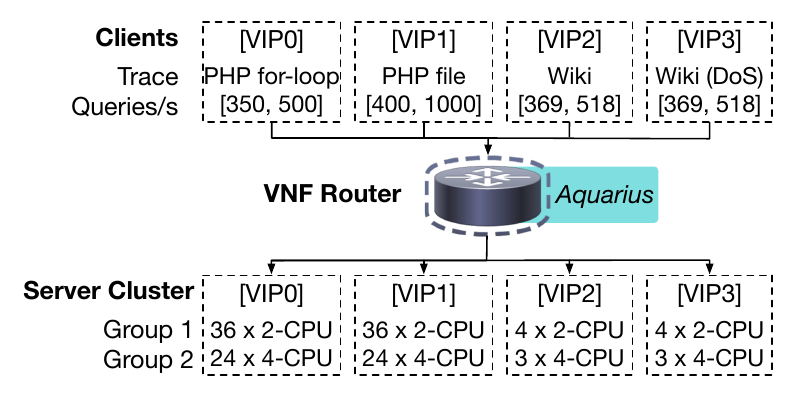}
	\vspace{-.1in}
	\caption{Network topology for traffic classification.}
	\label{fig:app-class-topo}
	\vskip -.1in
\end{figure}


\subsubsection{Task Description and Testbed Configuration}
\label{sec:application-classification-testbed}

This section shows the capability of \Albatross\ to collect reliable features and conduct traffic classification with unsupervised ML algorithms.
A testbed is implemented using Kernel-based Virtual Machine (KVM), where a virtual router embedded with \Albatross\ forwards different types of traffic to $4$ VIPs (Fig.~\ref{fig:app-class-topo}).
In VIP0, a simple PHP \texttt{for}-loop script on each server takes requests for given number of iterations and replies with proportional sizes.
The flow duration and size follow an exponential distribution as in~\cite{facebook-dc-traffic}.
In VIP1, static files of different sizes are served on each server\footnote{The sizes of files are $100$KB, $200$KB, $500$KB, $750$KB, $1$MB, $2$MB, and $5$MB. $50$ files are generated for each size.} as in~\cite{lbas-2020}, to represent IO-bound applications.
In VIP2 and VIP3, each application server is an independent replica of an Apache HTTP server that serves Wikipedia databases.
Two samples of $600$s duration are extracted and replayed from a real-world $24$-hour replay~\cite{wiki_traces}.
In VIP3, an additional $5000$ queries per second SYN flooding traffic is applied to simulate a DoS attack.

\begin{figure}[t]
	\centering
	\begin{subfigure}{.48\columnwidth}
		\centering
		\includegraphics[width=\columnwidth]{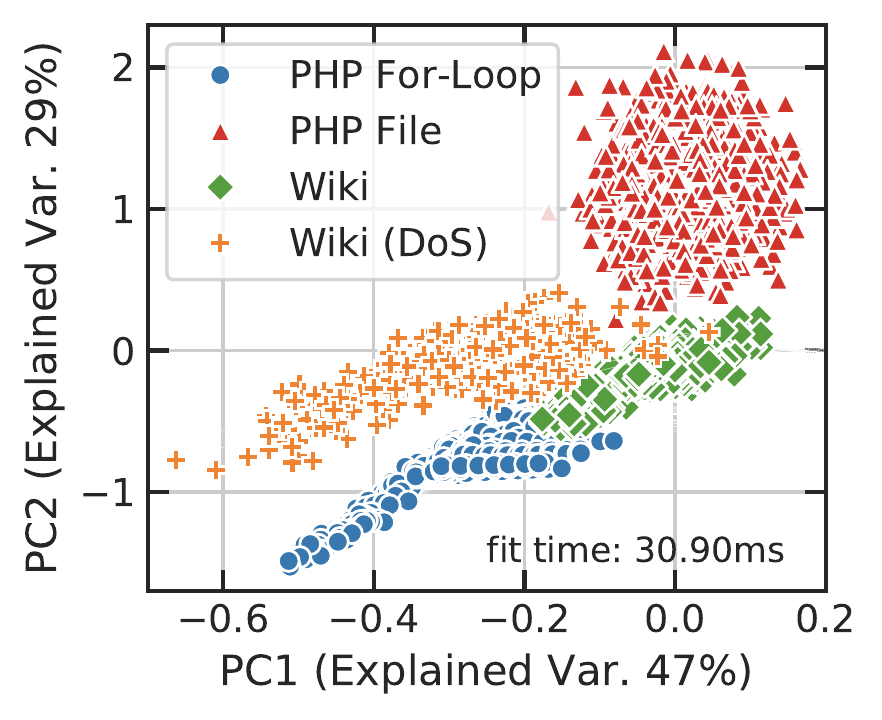}
		\vskip -0.05in
		\caption{PCA clusters (all features).}
		\label{fig:app-class-pca-cluster-full}
	\end{subfigure}
	\begin{subfigure}{.49\columnwidth}
		\centering
		\includegraphics[width=\columnwidth]{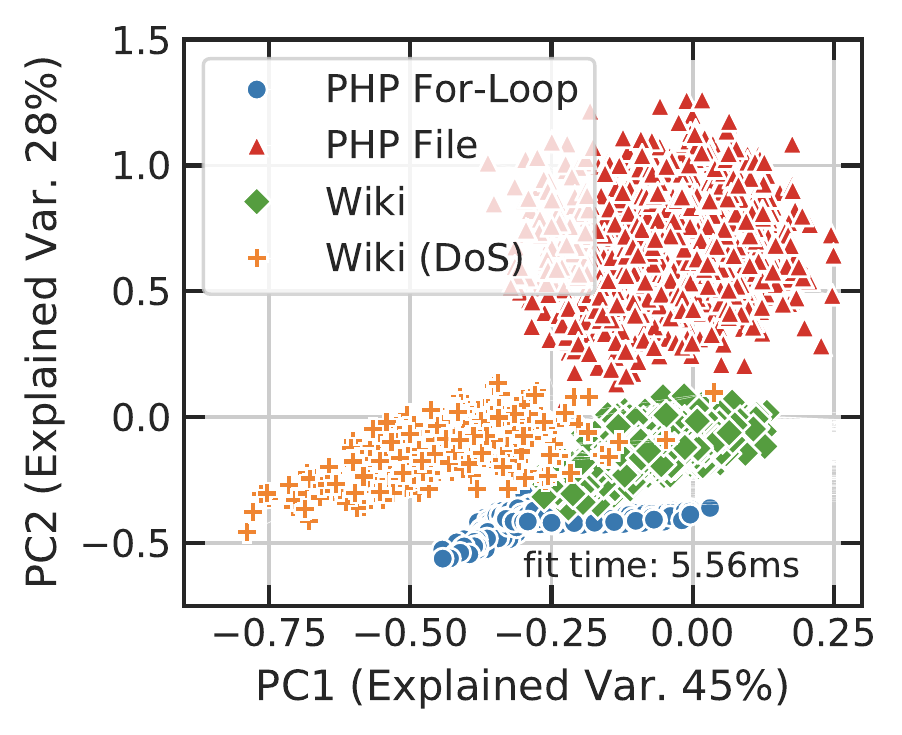}
		\vskip -0.05in
		\caption{PCA clusters ($25$ features).}
		\label{fig:app-class-pca-cluster-partial}
	\end{subfigure}
	\vskip -0.05in
	\caption{PCA analysis and $2$D visualisation.}
	\label{fig:app-class-pca-cluster}
	\vskip -0.2in
\end{figure}

\subsubsection{Principal Component Analysis (PCA) and Clustering}
\label{sec:application-classification-selection}

PCA is conducted to visualise the $4$ clusters for the $4$ different types of network traces in a $2$D projection (Fig.~\ref{fig:app-class-pca-cluster-full}).
Among the $4$ traces, PHP \texttt{for}-loop is pure CPU-bound and PHP file is pure IO-bound.
The \texttt{Wiki} trace consists of both queries for SQL database (CPU-bound) and static files (IO-bound), thus its cluster is located between the former $2$ traces.
The \texttt{Wiki} trace under DoS attack, however, can be clearly noticed as an independent cluster.
More features give multi-dimensional observations, yet at the cost of higher computation and memory overhead.
PCA helps reduce feature dimensionality from $73$ to $25$, while preserving data representation.
As depicted in Fig.~\ref{fig:app-class-pca-cluster-partial}, using $25$ features still gives clear clustering results, yet it reduces data processing time from $30.90$ms to $5.56$ms.

\subsubsection{Unsupervised Learning}
\label{sec:application-classification-unsupervised}

\begin{figure}[t]
	\begin{center}
		\centerline{\includegraphics[width=\columnwidth]{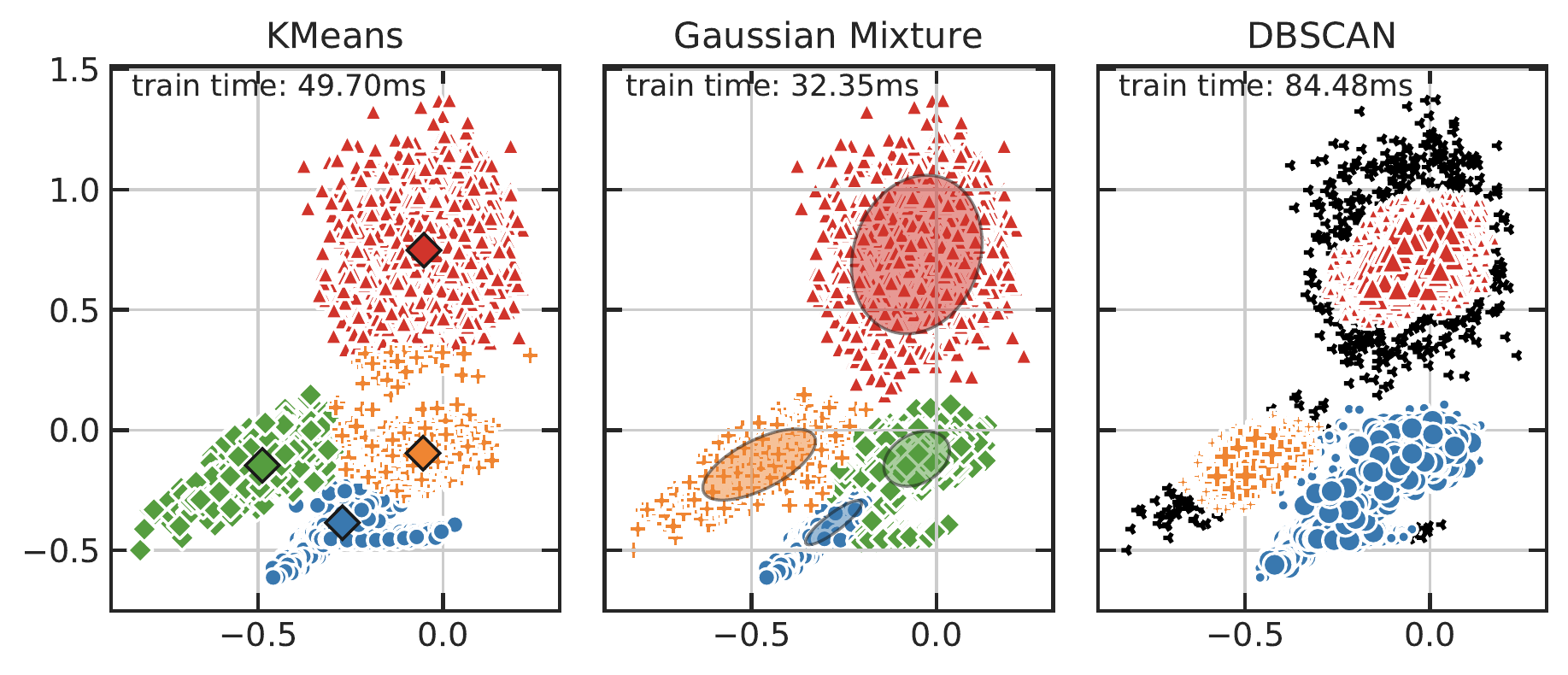}}
		\vskip -0.1in
		\caption{Unsupervised clustering using $25$ features.}
		\label{fig:app-class-pca-cluster-predict}
	\end{center}
	\vskip -0.2in
\end{figure}

As depicted in Fig.~\ref{fig:app-class-pca-cluster-predict}, when applying unsupervised learning algorithms, K-Means and Gaussian Misture are able to generate clusters similar to the ground truth, while they require the number of expected clusters ($4$) as input.
Gaussian Mixture model has the shortest fit time and can be an interesting candidate for online traffic classification.
In case where the number of clusters is not known \textit{a priori}, DBSCAN~\cite{schubert2017dbscan} can distinguish the potential security threat, based only on a predefined distance of $0.1$.

\begin{table}[t]
    \resizebox{\columnwidth}{!}{ 
        \begin{tabular}{|ll|r|r|r|r|}
        \hline
        \multicolumn{2}{|l|}{Configuration}                                                         & $0$ Feature & $11$ Features & $73$ Features & PCAP \\ \hline
        \multicolumn{1}{|l|}{\multirow{3}{*}{\rotatebox[origin=c]{90}{\begin{tabular}[c]{@{}c@{}}First\\ Packet\end{tabular}}}} & CPU Cycles                 & $938.232$      & $1635.838$     & $2609.019$  & $1295.284$   \\ \cline{2-6} 
        \multicolumn{1}{|l|}{}                              & Delay ($\mu$s) & $0.361$        & $0.629$        & $1.003$ 	& $0.498$         \\ \cline{2-6} 
        \multicolumn{1}{|l|}{}                              & Difference & $1.000\times$        & $1.744\times$        & $2.781\times$  & $1.381\times$      \\ \hline
        \multicolumn{1}{|l|}{\multirow{3}{*}{\rotatebox[origin=c]{90}{\begin{tabular}[c]{@{}c@{}}Data\\ Packet\end{tabular}}}}  & CPU Cycles                 & $576.357$      & $1583.798$     & $2602.684$  & $885.041$    \\ \cline{2-6} 
        \multicolumn{1}{|l|}{}                              & Delay ($\mu$s) & $0.222$        & $0.609$        & $1.001$ 	& $0.340$        \\ \cline{2-6} 
        \multicolumn{1}{|l|}{}                              & Difference          & $1.000\times$        & $2.748\times$        & $4.516\times$ 	& $1.536\times$        \\ \hline
        \multicolumn{2}{|l|}{CPU Usage (\%)}                                                         & $26.687$ & $40.858$ & $49.716$ & $31.480$ \\ \hline
        \multicolumn{2}{|l|}{CPU Difference}                                                         & $1.000\times$ & $1.376\times$ & $1.675\times$ & $1.060\times$ \\ \hline
        \multicolumn{2}{|l|}{RAM Usage (GiB)}                                                         & $2.652$ & $2.719$ & $2.744$ & $3.305$ \\ \hline
        \multicolumn{2}{|l|}{RAM Difference}                                                         & $1.000\times$ & $1.025\times$ & $1.034\times$ & $1.246\times$ \\ \hline
        \end{tabular}
    }
    \caption{Per-packet processing overhead (on $2.6$GHz CPU) and system resource consumptions (avg.) comparison.}
	\label{tab:app-class-overhead}
	\vskip -0.2in
\end{table}

\begin{figure}[t]
	\centering
	\begin{subfigure}{\columnwidth}
		\centering
		\includegraphics[width=\columnwidth]{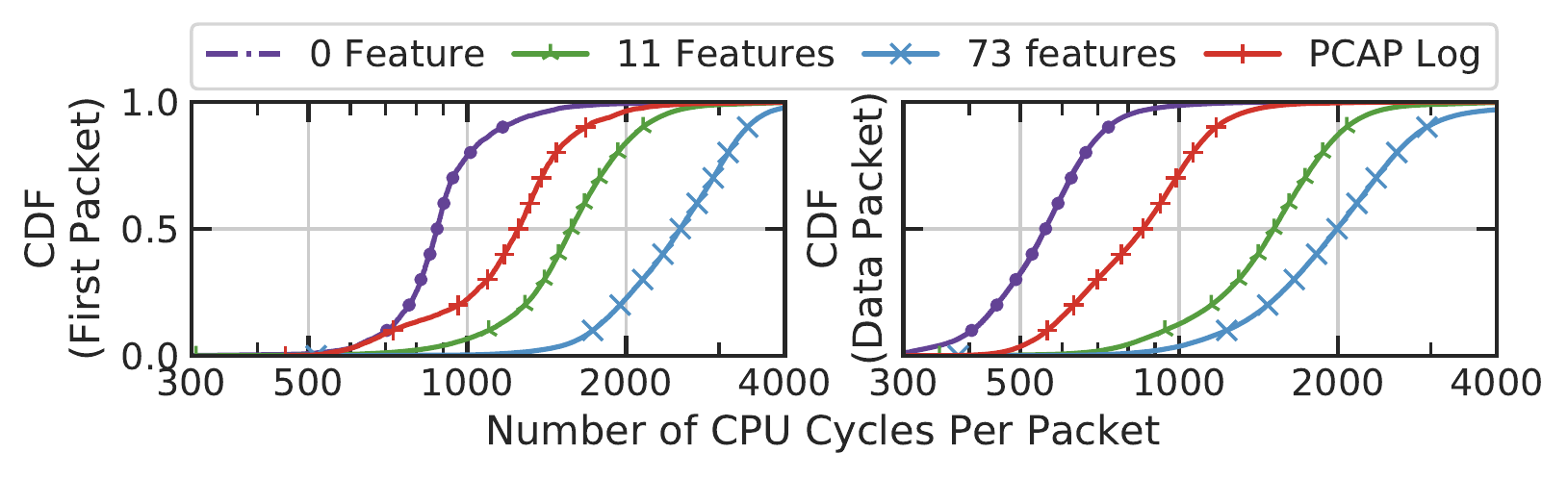}
		\vskip -0.05in
		\caption{Per-packet processing latency comparison.}
		\label{fig:application-classification-overhead-latency}
	\end{subfigure}
	\hspace{.35in}
	\begin{subfigure}{\columnwidth}
		\centering
		\includegraphics[width=\columnwidth]{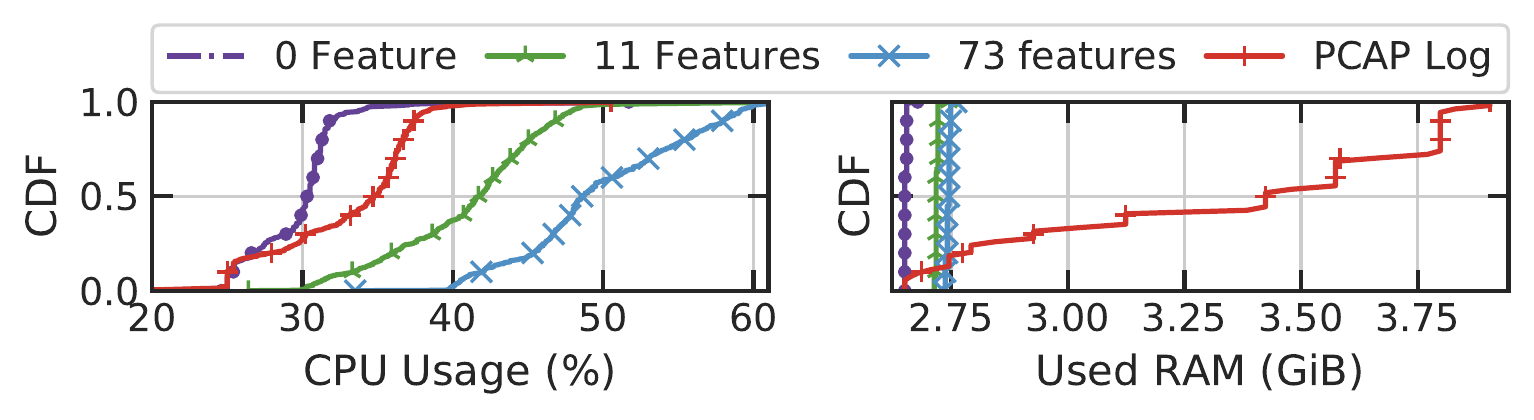}
		\vskip -0.05in
		\caption{System resource consumption.}
		\label{fig:application-classification-overhead-resource}
	\end{subfigure}
	\vspace{.1in}
	\begin{subfigure}{\columnwidth}
		\centering
		\includegraphics[width=\columnwidth]{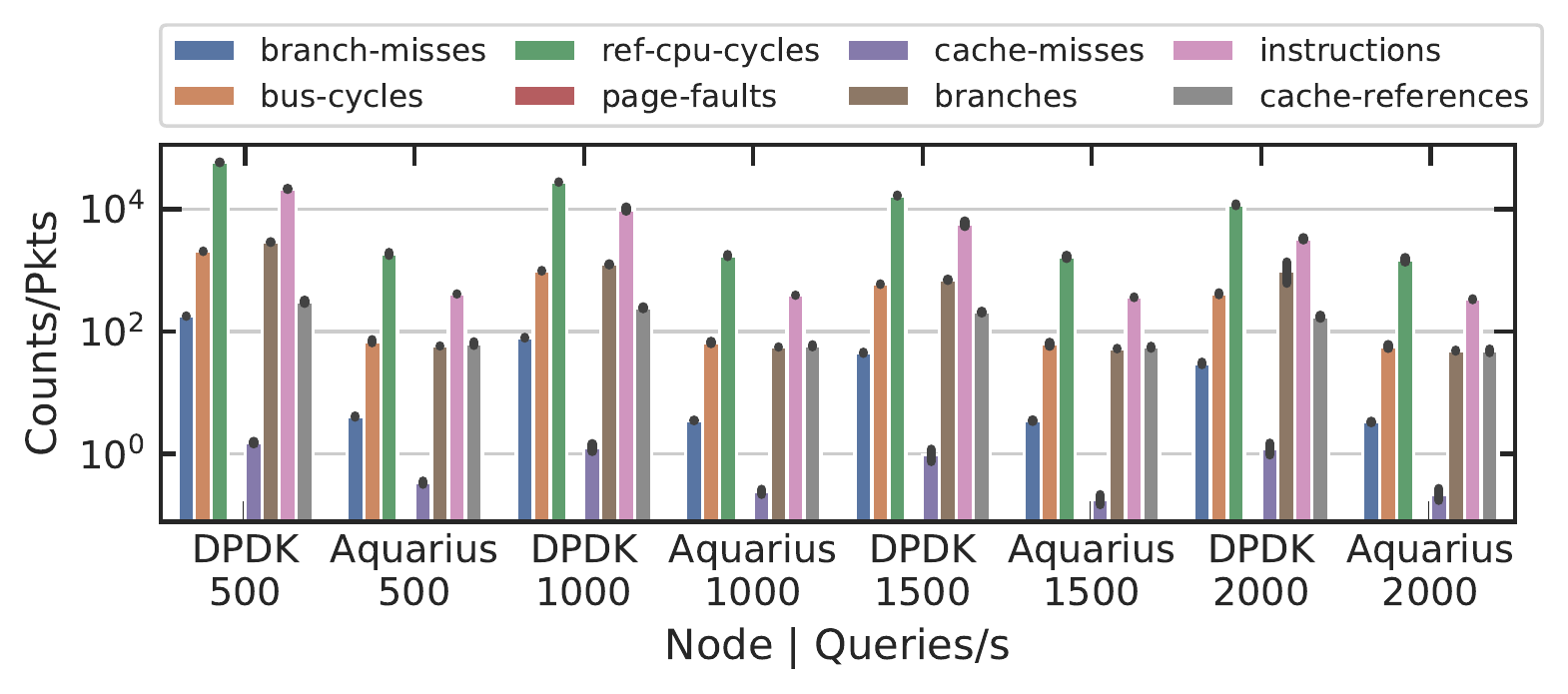}
		\vskip -0.1in
		\caption{System performance metric comparison}
		\label{fig:micro-flow-table-sys-perf}
	\end{subfigure}
	\vskip -0.1in
	\caption{\Albatross\ feature collection overhead.}
	\label{fig:application-classification-overhead}
	\vskip -0.1in
\end{figure}

\subsubsection{Overhead Analysis}
\label{sec:application-classification-overhead}

To study the feature collection overhead, \Albatross\ is compared with a vanilla router which collects $0$ feature and a router logging packet information in the memory using \texttt{pcap}.
Under $500$ queries/s PHP \texttt{for}-loop traffic towards a $176$-CPU server cluster, when collecting $11$ features or collecting all $73$ features, \Albatross\ incurs different overhead (Table~\ref{tab:app-class-overhead} and Fig.~\ref{fig:application-classification-overhead-latency}).
On a $2.6$GHz CPU, the additional per-packet processing delays are trivial comparing with the typical round trip time (higher than $200\mu$s) between network equipments.
The mean CPU usage of \Albatross\ is $1.376\times$ and $1.675\times$ higher than the vanilla router when collecting $11$ and $73$ features respectively (Fig.~\ref{fig:application-classification-overhead-resource}).
As expected, log-based feature collection mechanism does not scale in terms of memory consumption\footnote{The results can be machine-dependent. This \publicationtype\ aims at showing the order of magnitudes, rather than providing a precise quantification.}.
As depicted in Fig.~\ref{fig:micro-flow-table-sys-perf}, the system performance of \Albatross\ shows advantages over RX interface driven by DPDK on various metrics when using different traffic rates.

\textbf{Take-Away:}
\Albatross\ gathers fine-grained and reliable datasets, which allow feature engineering and conducting in-depth data analysis.
Its fast and configurable design help achieve the right balance between visibility and performance.

\subsection{Resource Prediction and Auto-Scaling}
\label{sec:application-prediction}


To minimize operational cost while guaranteeing QoS, cloud operators~\cite{aws-elastic}) need to elastically provision server capacities.

\begin{figure}[t]
	\begin{center}
		\centerline{\includegraphics[width=.9\columnwidth]{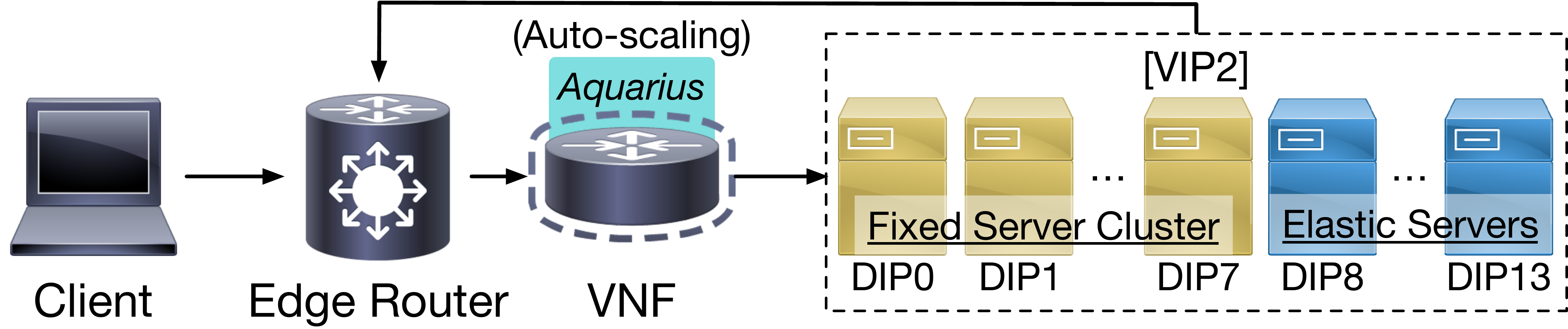}}
		\caption{Network topology for autoscaling system.}
		\label{fig:app-predict-topo}
	\end{center}
	\vskip -0.45in
\end{figure}


\subsubsection{Task Description and Testbed Configuration}
\label{sec:application-prediction-testbed}

This section shows the capability of \Albatross\ as a platform to adapt supervised ML algorithms to infer resource utilisation with no actively signaling.
$600$s samples extracted from the real-world $24$-hour Wikipedia trace are replayed on the network topology depicted in Fig.~\ref{fig:app-predict-topo}.
Workloads are randomly distributed among running servers ($2$-CPU each) by way of ECMP.
A learning task can be framed as predicting server load states (CPU usage) on each server with the same set of features as in Sec.~\ref{sec:application-classification}.
The predicted utilisation can be then used to plan and re-scale server cluster to guarantee QoS with reduced operational cost.
This task consists of $2$ steps--offline model training and online prediction.

\subsubsection{Offline Model Training}
\label{sec:application-prediction-offline}

To predict the resource utilisation of server clusters using networking features, $12$ widely used ML algorithms are selected to cover different families of ML algorithms, \eg sequential and non-sequential, parametric and non-parametric, linear and non-linear~\cite{fu2021use}.
The first $23$-hour samples are applied on $10$ servers to gather datasets for offline model training.
The distribution of the ground truth CPU usages in the training set covers the other two datasets so that the prediction task is feasible, yet the ML models have not seen the datasets for evaluations.
Based on the offline training performance evaluation, $1$ non-sequential model (linear regression) and $1$ sequential model (WaveNet~\cite{oord2016wavenet}) are selected to be applied for online auto-scaling\footnote{The whole comparison among all $12$ models and additional experimental results will be included supplementary details.}.

\alglanguage{pseudocode}
\begin{algorithm}[t]
\footnotesize
\caption{Auto-scaling Rule}
\label{alg:autoscale}
\begin{algorithmic}[1]
    \State $n\_servers\_min,n\_servers\_max \gets 8, 14$ \Comment{Server number range}
    \State $\mathbb{S} \gets$ Initial set of running servers
    \State $cpu\_lo,cpu\_hi \gets 0.7, 0.8$ \Comment{Desired CPU usage range}
    \For {each time step} \Comment{$\Delta t$ = $250$ms}
        \State $\delta \gets 0$ \Comment{Initialize server state counter}
        \State $y(\mathbb{S}) \gets$ CPU usage prediction of $16$ steps ahead
        \State $threshold \gets \lceil\frac{|\mathbb{S}|}{3}\rceil$ \Comment{Threshold that triggers scaling actions}
        \For {$s \in \mathbb{S}$}
            \If {$y(s) < cpu\_lo$}
                \State $\delta++$ \Comment{Increment $\delta$ if $s$ is under-loaded}
            \ElsIf {$y(s) > cpu\_hi$}
                \State $\delta--$ \Comment{Decrement $\delta$ if $s$ is over-loaded}
            \EndIf
        \EndFor
        \If {$\delta > threshold$ \textbf{and} $|\mathbb{S}| > n\_servers\_min$}
            \State $\mathbb{S} \gets downscale(\mathbb{S})$
			\State skip $8$ time steps \Comment{Cool-down period}
        \ElsIf {$\delta < -threshold$ \textbf{and} $|\mathbb{S}| < n\_servers\_max$}
            \State $\mathbb{S} \gets upscale(\mathbb{S})$
			\State skip $8$ time steps \Comment{Cool-down period}
        \EndIf
    \EndFor
\Statex
\end{algorithmic}
\vskip -0.05in
\end{algorithm}

\begin{figure}[t]
	\vskip -.2in
	\begin{center}
		\centerline{\includegraphics[width=\columnwidth]{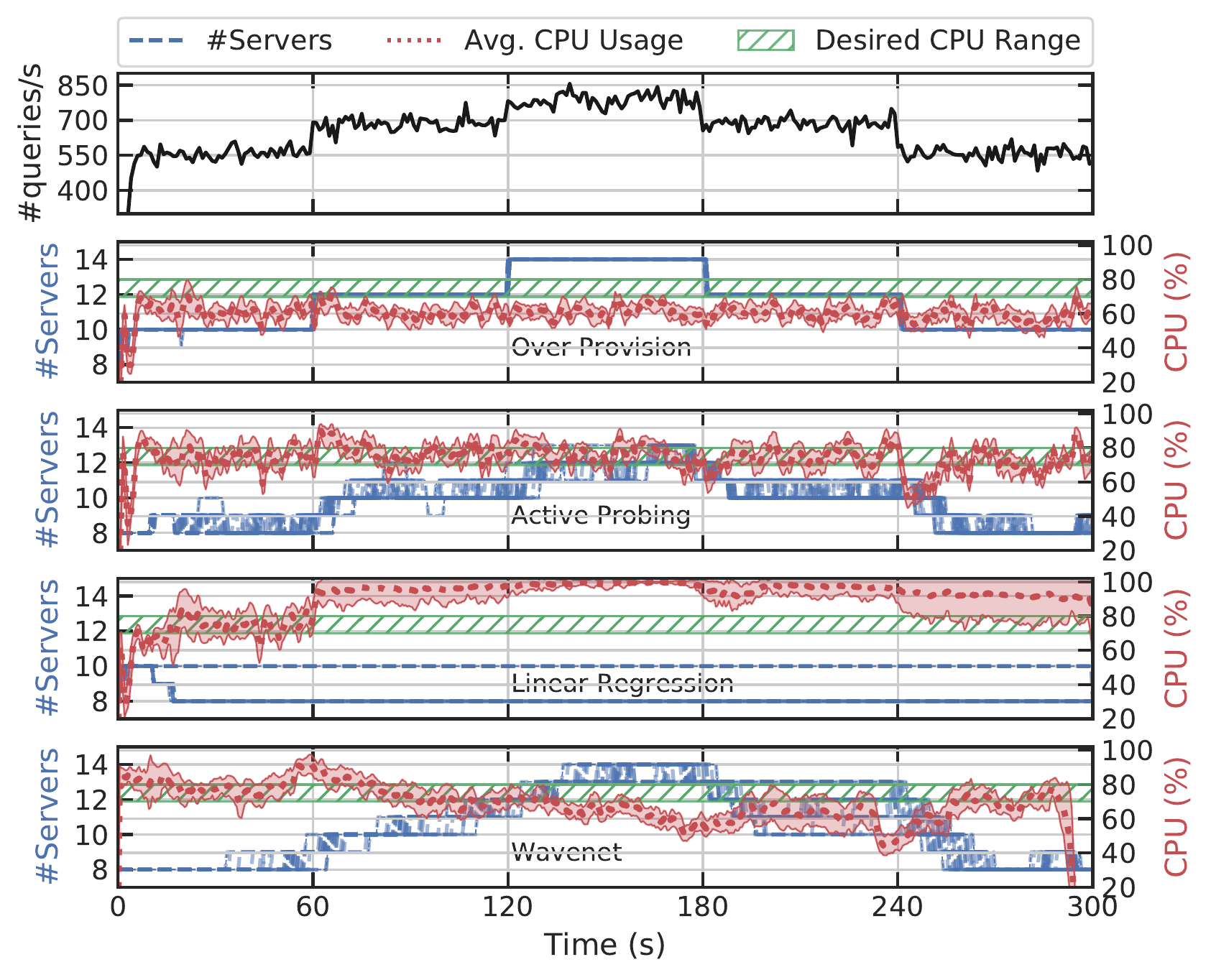}}
		\vskip -0.1in
		\caption{Comparison of online auto-scaling performance using different algorithms.
        The (discrete) numbers of running servers are plotted for each run in dashed lines, while CPU usage is summarised as avg. $\pm$ stddev across $30$ runs.}
		\label{fig:app-predict-timeline}
	\end{center}
	\vskip -0.35in
\end{figure}

\subsubsection{Online Auto-Scaling}
\label{sec:application-prediction-online}

To test the online performance of offline-trained ML models, a $300$s Wikipedia replay trace sample of the last hour (unseen by the ML models) is synthesized to have scheduled changing traffic rates every $60$s.
Based on the CPU usage predictions of running servers $y(\mathbb{S})$, a simple heuristic (Alg.~\ref{alg:autoscale}), which approximate the threshold-based autoscaling policy as in~\cite{aws-elastic}, is proposed to keep the CPU usage of $\frac{2}{3}$ servers within the desired range ($70\sim80\%$).
Using the same counter $\Delta$ for over- and under-loaded servers reduces the variance induced by imbalanced workload distributions.
As a comparison to~\cite{aws-elastic}, a threshold-based active probing mechanism is implemented, whose predicted CPU usage for running servers $y(\mathbb{S})$ come from periodic polling.
An ``oracle'' benchmark is implemented to over-provision the number of servers proportional to the scheduled traffic rates.

\begin{figure}[t]
	\centering
	\begin{subfigure}{\columnwidth}
		\centering
		\includegraphics[width=.9\columnwidth]{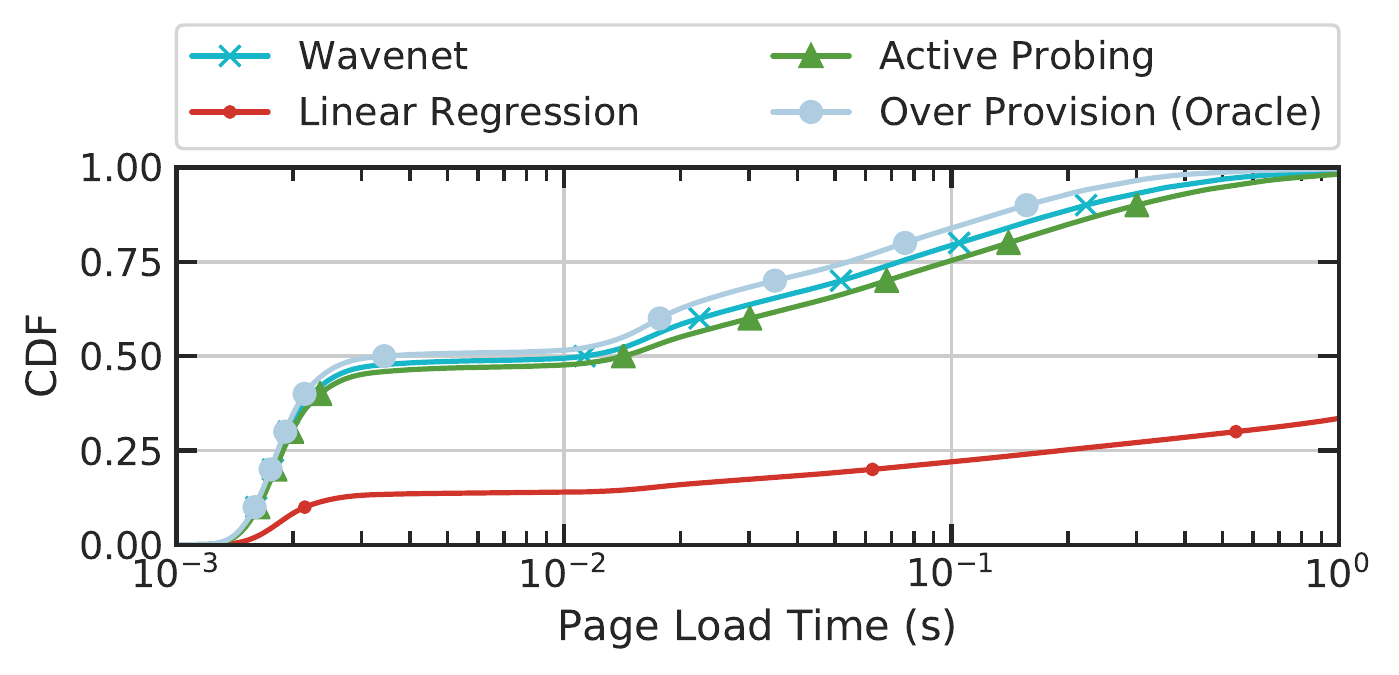}
		\vskip -0.1in
		\caption{QoS.}
		\label{fig:app-predict-qos}
	\end{subfigure}
	\begin{subfigure}{\columnwidth}
		\centering
		\includegraphics[width=.9\columnwidth]{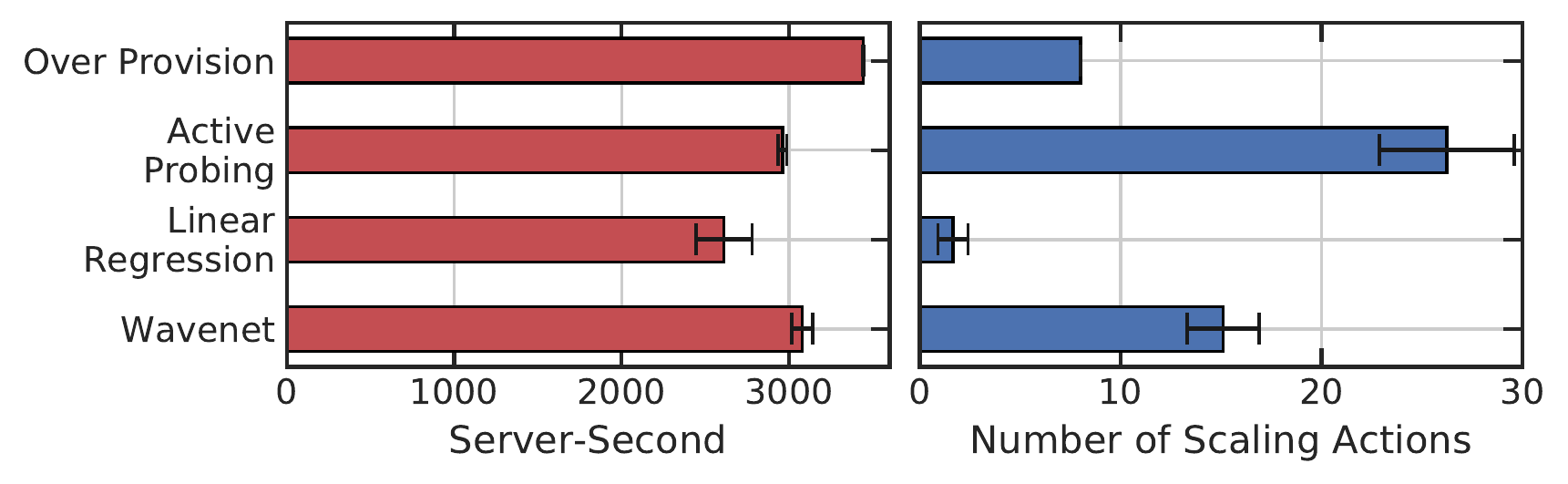}
		\vskip -0.1in
		\caption{Operational cost and complexity.}
		\label{fig:app-predict-cost}
	\end{subfigure}
	\vskip -0.05in
	\caption{Trade-off between QoS and cost using different autoscaling mechanisms.}
	\label{fig:app-predict-trade-off}
	\vskip -0.2in
\end{figure}

\begin{figure}[t]
	\begin{center}
		\centerline{\includegraphics[width=.9\columnwidth]{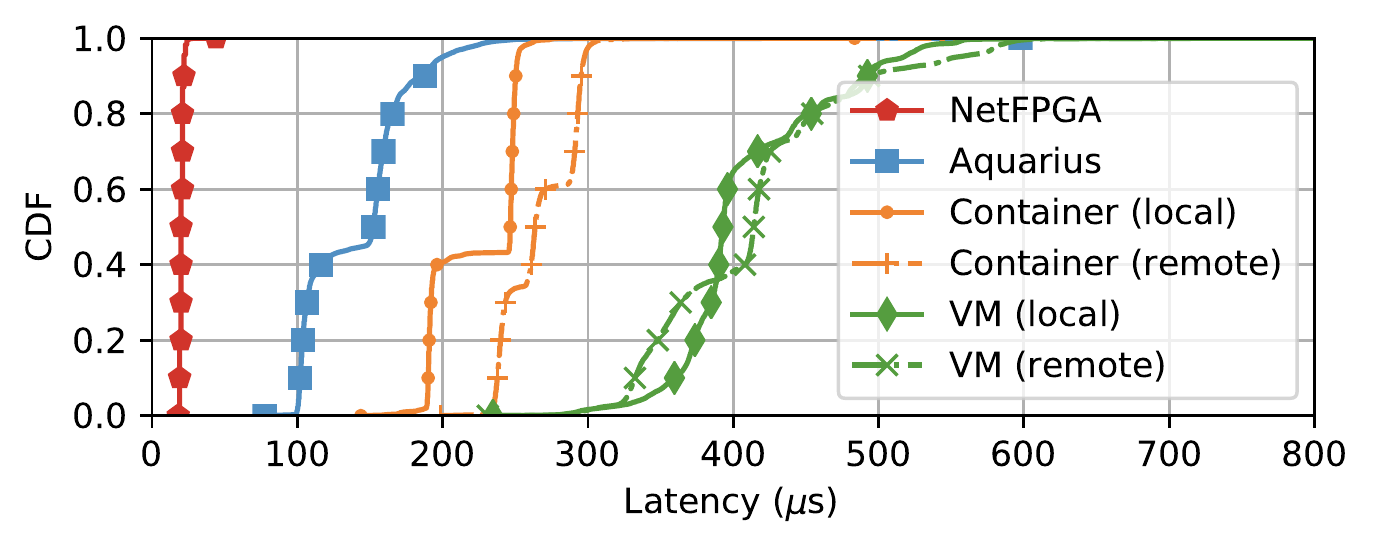}}
		\vskip -0.1in
		\caption{Feature collection latency comparison between \Albatross\ and active probing techniques.}
		\label{fig:online-probe}
	\end{center}
	\vskip -0.4in
\end{figure}

\subsubsection{Results}
\label{sec:application-prediction-results}

As depicted in Fig.~\ref{fig:app-predict-timeline}, 
active probing keeps the average CPU usage within the desired range, however, it requires frequent scaling events and leads to oscillating CPU usage with high variance.
Linear regression is simple yet not robust when applied for an online auto-scaling system.
Its under-estimated server load states lead to over-loaded servers.
WaveNet takes sequential features as input and is more robust when applied online.
It keeps the average CPU usage close to the desired range with less oscillations.

As depicted in Fig.~\ref{fig:app-predict-trade-off}, WaveNet is able to provide better QoS than active probing--$78.37$ms less page load time ($26.04\%$) at $90$th percentile and $35.70$ms less ($30.45\%$) on average--with
$3.99\%$ additional server-second cost, and $42.44\%$ less scaling events.
When over-provisioning the server cluster, the page load time is lower than using WaveNet by 
$67.13$ms at $90$th percentile and 
$28.55$ms average, though it requires $11.86\%$ more server-second operational cost than WaveNet.

\Albatross\ parses features stored in the local shared memory with no control messages, achieving more than $94.18\mu$s less median latency than typical VM- and container-based probing mechanisms (Fig.~\ref{fig:online-probe}).

\textbf{Take-Away:}
\Albatross\ enables agile development and online deployment of learning algorithms to improve network performance.
It makes features quickly accessible while saving management bandwidth for data transmission.



\subsection{Traffic Optimisation and Load Balancing}
\label{sec:application-lb}

As a key component in cloud DCs, Layer-$4$ load balancers (LBs) distribute workloads across servers to provide scalable services.
Yet, in-production LBs requires human intervention which can lead to server weights mis-configurations.

\begin{figure}[t]
	\centering
	\begin{subfigure}{\columnwidth}
		\centering
		\includegraphics[width=\columnwidth]{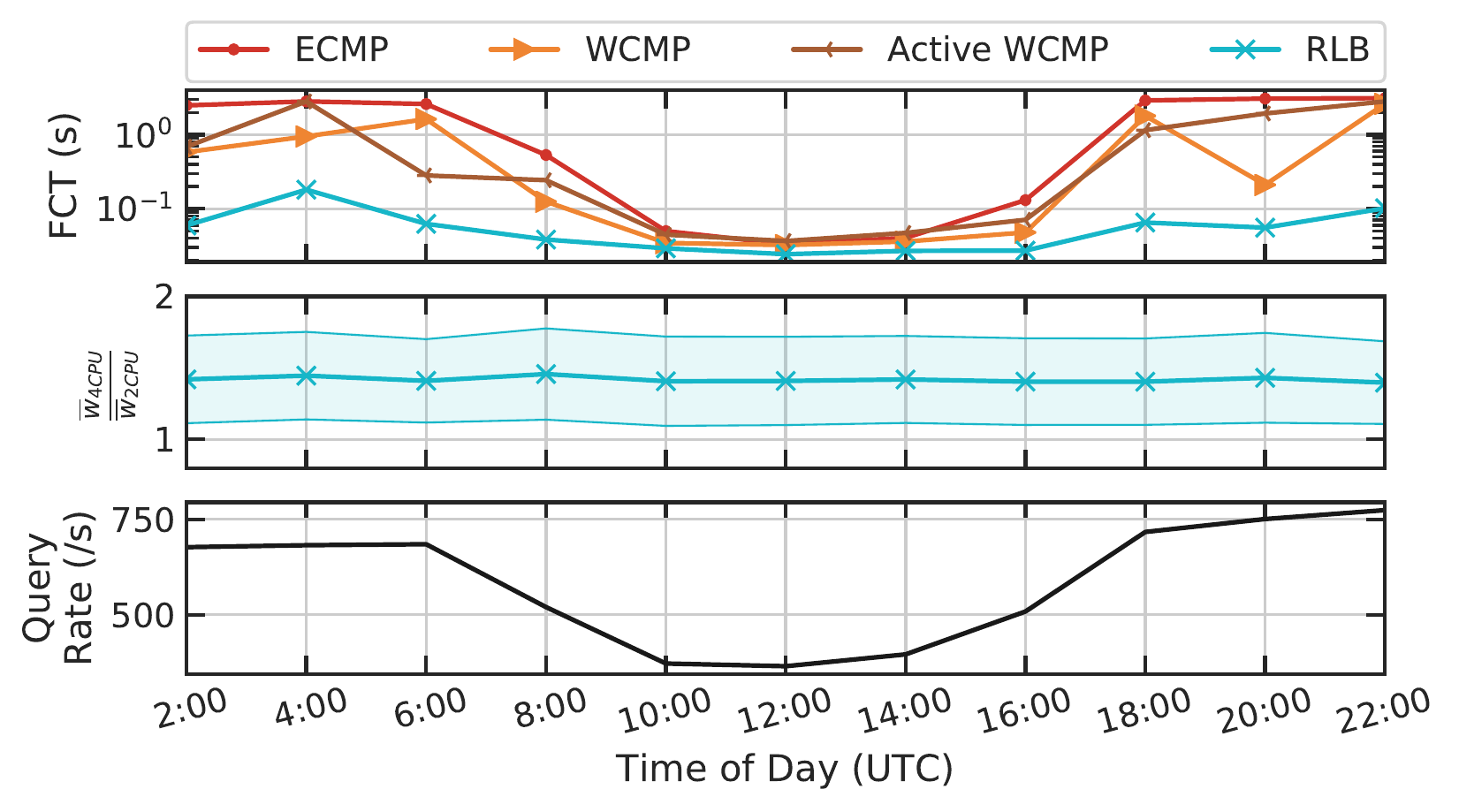}
		\vskip -0.1in
		\caption{Mean FCTs (top), ratio between weights assigned to the $2$ groups of servers by RLB (middle), and traffic rates (bottom).}
		\label{fig:app-lb-wiki-fct-timeline}
	\end{subfigure}
	\begin{subfigure}{\columnwidth}
		\centering
		\includegraphics[width=\columnwidth]{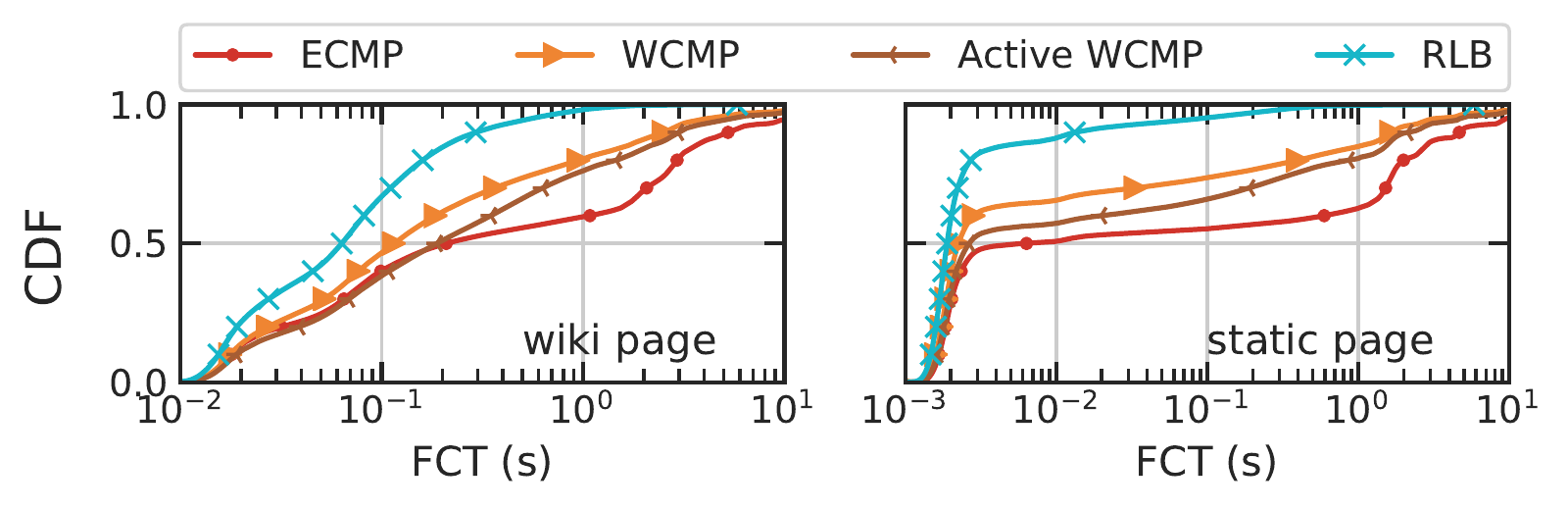}
		\vskip -0.1in
		\caption{Peak-hour (query rate higher than $500$/s) FCT distribution.}
		\label{fig:app-lb-wiki-fct-peak}
	\end{subfigure}
	\vskip -0.05in
	\caption{Wikipedia trace replayed using different LBs.}
	\label{fig:app-lb-wiki-fct}
	\vskip -0.2in
\end{figure}

\subsubsection{Task Description and Testbed Configuration}
\label{sec:application-lb-description}

This section shows that \Albatross\ can apply RL algorithms to self-configure server weights and optimise load balancing performance with no-prior knowledge of the system.
The configuration of VIP2 (Fig.~\ref{fig:app-class-topo}) is applied--replaying the \texttt{Wiki} trace and load balancing on $2$ groups of servers of different processing capacities.
The task is to extract and infer server processing capacity information from networking features and make informed load balancing decisions.
$3$ benchmark LB algorithms are implemented--(i) ECMP~\cite{silkroad2017}, (ii) statically configured WCMP~\cite{ananta2013}, and (iii) active WCMP~\cite{lbas-2020} based on polled server job queue lengths.
Both ECMP and WCMP are based on the kernel-bypassing implementation of Maglev~\cite{maglev} in VPP, which is the state-of-the-art LB.

\subsubsection{RL Algorithm}
\label{sec:application-lb-algorithm}

This \publicationtype\ implements and evaluate RLB~\cite{yao2021reinforced}--implemented and evaluated in simulators--in a realistic testbed using \Albatross.
With \Albatross, RLB (i) counts ongoing flows $\tilde{l}_i$ on servers and (ii) asynchronously updates (every $250$ms) server weights $\tilde{w}_i$ derived from flow durations $\tau_i$ sampled in reservoir buffers.
On receipt of new requests, RLB assigns servers based on $\argmin_{i} \frac{\tilde{l}_{i} + 1}{\tilde{w}_{i}}$, which prioritizes servers with higher processing speed and shorter queue lengths.
Different from~\cite{yao2021reinforced}, which uses actively probed ground truth information, this \publicationtype\ derives the reward from locally observed features (FCT) collected by \Albatross, which requires no active signalling between LBs and servers.
RLB is trained using the first hour of \texttt{Wiki} trace sample for $20$ episodes.
The trained RLB model is then tested on unseen traffic and compared with other LB algorithms.

\subsubsection{Results}
\label{sec:application-lb-result}

\begin{figure}[t]
	\begin{center}
		\centerline{\includegraphics[width=\columnwidth]{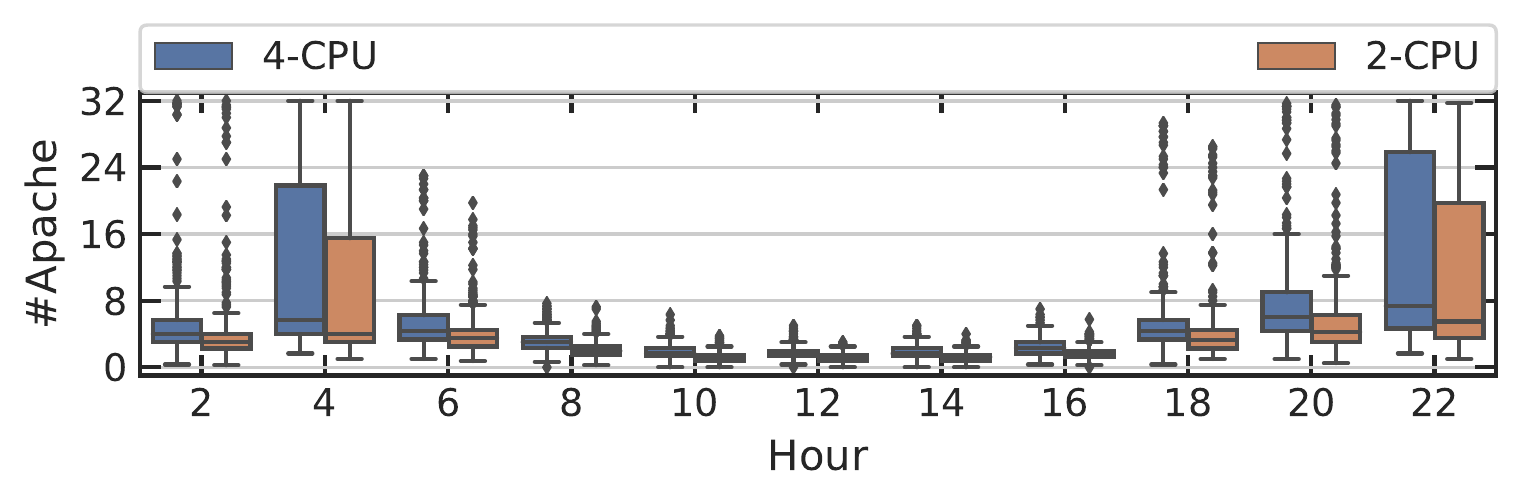}}
		\vskip -0.1in
		\caption{Query distribution (number of busy Apache threads) on $2$ groups of application servers.}
		\label{fig:app-lb-wiki-apache}
	\end{center}
	\vskip -0.4in
\end{figure}

As depicted in Fig.~\ref{fig:app-lb-wiki-fct-timeline}, during off-peak hours, servers are under-utilised and all algorithms show similar performances in terms of QoS.
As traffic rates grow, RLB achieves lower FCT for both static pages and Wikipedia pages when compared with other LB algorithms (Fig.~\ref{fig:app-lb-wiki-fct-peak}), since RLB is able to dynamically adjust server weights.
As depicted in Fig.~\ref{fig:app-lb-wiki-fct-peak}, for Wikipedia pages the $90$-th percentile FCT of RLB ($0.292$s) is $18.15$x shorter than Maglev ECMP and $8.56$x shorter than Maglev WCMP, and the $95$-th percentile FCT of RLB ($0.552$s) is $17.82$x shorter than Maglev ECMP and $7.82$x shorter than Maglev WCMP.
For static pages, the $90$-th percentile FCT of RLB ($0.013$s) is $351.15$x shorter than ECMP and $124.43$x shorter than WCMP, and the $95$-th percentile FCT of RLB ($0.088$s) is $109.09$x shorter than ECMP and $37.50$x shorter than WCMP.
RLB is trained to learn server processing speed differences and assigns higher weights, thus more queries, to more powerful servers (Fig.~\ref{fig:app-lb-wiki-apache}).
When using RLB, $4$-CPU servers handle respectively $1.258\times$ and $1.523\times$ more tasks than $2$-CPU servers under $676.92$ and $372.01$ queries/s traffic.

\subsubsection{Overhead Analysis}
\label{sec:application-lb-overhead}

\begin{figure}[t]
	\centering
	\begin{subfigure}{\columnwidth}
		\centering
		\includegraphics[width=\columnwidth]{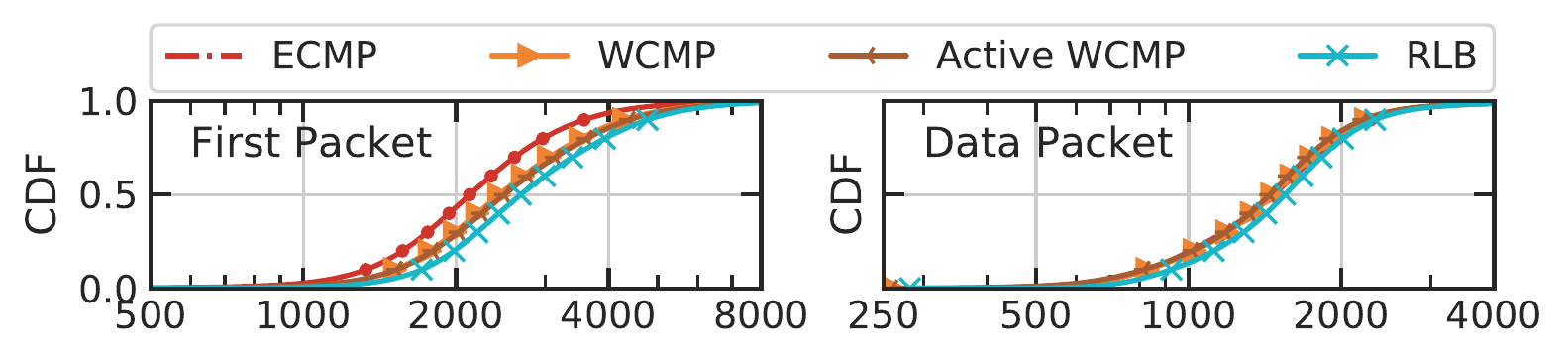}
		\vskip -0.05in
		\caption{Per-packet processing latency comparison.}
		\label{fig:app-lb-overhead-latency}
	\end{subfigure}
	\hspace{.35in}
	\begin{subfigure}{\columnwidth}
		\centering
		\includegraphics[width=\columnwidth]{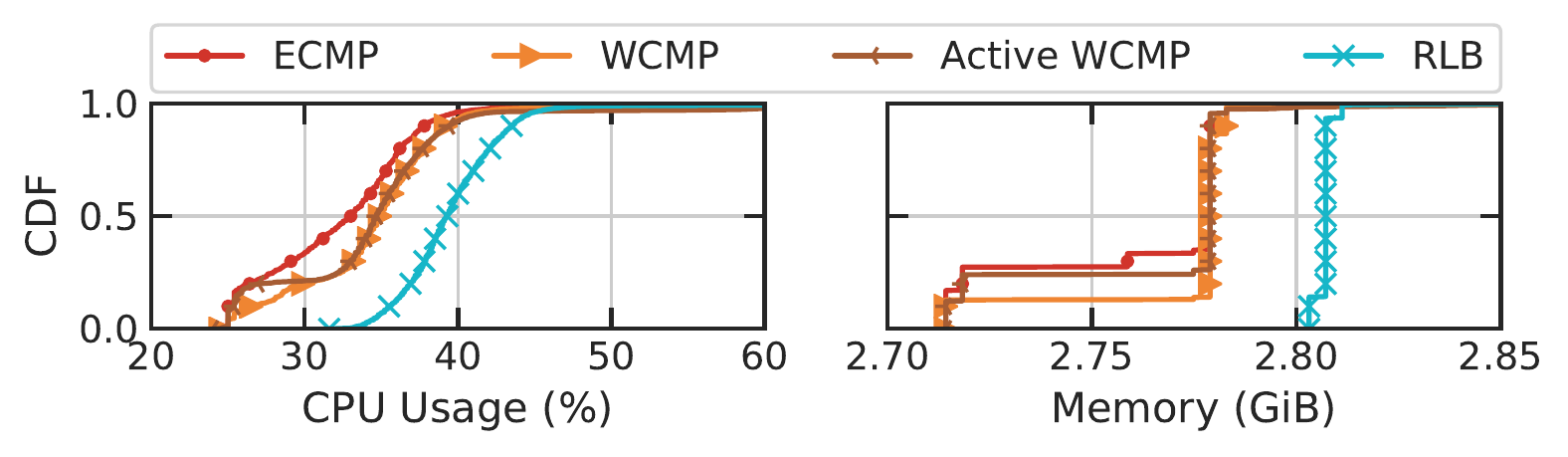}
		\vskip -0.05in
		\caption{System resource consumption.}
		\label{fig:app-lb-overhead-resource}
	\end{subfigure}
	\vskip -0.05in
	\caption{Overhead comparisons.}
	\vskip -0.2in
	\label{fig:app-lb-overhead}
\end{figure}

As depicted in Fig.~\ref{fig:app-lb-overhead-latency}, throughout all test runs, RLB consume on average $692.89$ more CPU cycles ($0.26\mu$s on $2.6$GHz CPU) than ECMP, as it computes and compares the server scores when making load balancing decisions.
Fig.~\ref{fig:app-lb-overhead-resource} depicts CPU and memory consumptions of all LBs.
On average, RLB incurs $0.22\times$ additional CPU usage, and $45.99$MiB memory usage, and achieves $87.38\%$ throughput of ECMP.

\textbf{Take-Away:}
\Albatross\ enables closed-loop control (RL) to dynamically adapt to networking systems and optimise performance.
It empowers real-world deployment and evaluation of learning algorithms developed in simulated environments.

\section{Conclusion}
\label{sec:conclusion}

Networking features and system state information help VNFs make informed decisions, and intelligently manage and update networking policies in cloud DCs.
Actively collecting features and system state information entails substantial control signalling and management overhead, in particular in large-scale DC networks.
This \publicationtype\ has proposed \Albatross, a framework that collects, infers and supplies accurate networking state information with little additional processing latency, in a scalable buffer layout.
The \publicationtype\ has illustrated significant performance gains of using of \Albatross\ for various ML-based VNFs and evaluated experimentally the impact of \Albatross\ in the system performance.


\bibliographystyle{IEEEtran}
\bibliography{reference}

\end{document}